\documentclass[%
 reprint,
 amsmath,amssymb,
 aps,
pra,
]{revtex4-1}
\usepackage{amsmath}
\usepackage{upgreek}
\usepackage{graphicx}
\usepackage{dcolumn}
\usepackage{bm}
\usepackage{listings}

\begin{document}


\title[]{Experimental simultaneous read out of the real and imaginary parts of the weak
value}

\author{A. Hariri}
\author{D. Curic}%
\author{L. Giner}%

\author{J. S. Lundeen}
\email[]{jlundeen@uOttawa.ca}
\affiliation{Department of Physics and Centre for Research in Photonics, University of Ottawa, 25 Templeton Street, Ottawa, Ontario, Canada K1N 6N5}

\date{\today}

\begin{abstract}
The weak value, the average result of a weak measurement, has proven
useful for probing quantum and classical systems. Examples include
the amplification of small signals, investigating quantum paradoxes,
and elucidating fundamental quantum phenomena such as geometric phase.
A key characteristic of the weak value is that it can be complex,
in contrast to a standard expectation value. However, typically only
either the real or imaginary component of the weak value is determined
in a given experimental setup. Weak measurements can be used to, in
a sense, simultaneously measure non-commuting observables. This principle
was used in the direct measurement of the quantum wavefunction. However,
the wavefunction's real and imaginary components, given by a weak
value, are determined in different setups or on separate ensembles
of systems, putting the procedure's directness in question. To address
these issues, we introduce and experimentally demonstrate a general
method to simultaneously read out both components of the weak value
in a single experimental apparatus. In particular, we directly measure
the polarization state of an ensemble of photons using weak measurement.
With our method, each photon contributes to both the real and imaginary
parts of the weak-value average. On a fundamental level, this suggests
that the full complex weak value is a characteristic of each photon
measured. 
\end{abstract}
\pacs{}

\keywords{Weak value, von Neumann measurement, weak measurement, quantum measurement}
\maketitle
\section{Introduction}

Weak values and weak measurement have attracted a considerable amount
of interest in recent years \cite{dressel2014colloquium}. Weak values
were introduced in 1988 \cite{Aharonov1988,Aharonov1989} as the average
result of a gently probing measurement (i.e., a `weak measurement')
of an quantum observable for an input quantum state and followed by
a projective measurement. They have been used to investigate quantum
paradoxes, such as the three-box \cite{Aharonov1991,resch2004experimental},
the Cheshire Cat \cite{Aharonov2013,Denkmayr2014}, and Hardy's paradoxes
\cite{aharonov2002revisiting,Lundeen2009a}. Weak values are deeply
connected to other fundamental and uniquely quantum phenomena, such
as geometric phase \cite{sjoqvist2006geometric}, time-reversal symmetry
violation \cite{Aharonov1964,Bednorz2013,Curic2018}, Bayesian quantum
estimation \cite{Johansen2004a}, and non-contextuality \cite{tollaksen2007pre,dressel2010contextual,pusey2014anomalous}.
They have also found application in the field of metrology. In a technique
called `weak-value amplification', the weak value can become much
larger than the standard expectation value of an observable, thereby
amplifying the associated measurement signal \cite{Aharonov1988}.
This has been used to measure small shifts in quantities such as phase,
time, frequency, angle, and temperature \cite{hosten2008observation,starling2009optimizing,starling2010continuous,brunner2010measuring,egan2012weak}.

Most relevant to this work, weak measurement has been used to directly
measure the quantum wavefunction of a system \cite{lundeen2011direct,Lundeen2012,Bamber2014,Thekkadath2016}.
This procedure weakly measures one variable (e.g., position $x$)
and then strongly measures the complementary variable (e.g., momentum
$p$). The real and imaginary parts of the wavefunction $\psi(x)$
appear directly on the measurement apparatus as the real and imaginary
parts of the weak value. In this way, an unknown input quantum state
can be determined. And, in contrast to quantum state tomography, this
can be accomplished without the need for a complicated mathematical reconstruction such as an inversion or fitting.
Nevertheless, there has been a degree of indirectness in almost all
experiments that measure both the real and imaginary parts of the
weak value. In particular, each part is measured by averaging over
a separate sub-ensemble. This happens in two possible ways: 1. A different
apparatus is used to measure each of the two parts. 2. A single apparatus
randomly chooses whether it is the real or imaginary part that is
measured in a given trial. In either case, the real and imaginary
parts of the wavefunction are determined separately, which diminishes
the purported directness of the technique.

Another motivation for this work is to establish whether the weak
value is fundamental to a physical system. The experiments described
above suggest that measurements of the real and the imaginary parts
of the weak value are mutually exclusive. It may be fundamentally
impossible to measure them simultaneously, much like the Heisenberg
Uncertainty Principle forbids the simultaneous precise measurement
of complementary observables. If true, it may be incorrect to consider
both the real and imaginary parts of the weak value as the average
result of weak measurement or as simultaneous properties of the measured
system. This would contradict particular interpretations of quantum
physics that take weak values as deterministic (i.e., `real' or ontological)
properties of a system \cite{Aharonov1991,Vaidman1996a,Hofmann2012a}.

Recently, an experiment demonstrated the simultaneous read out of
the real and imaginary parts of the weak value using orbital angular
momentum states of light \cite{Kobayashi2014}. However, it was unclear
whether orbital angular momentum was inherently necessary or whether
the technique was more general. In this work, we show that the technique
in Ref. \cite{Kobayashi2014} can be generalized to a broad class
of weak measurement implementations by performing a sequence of two
separate weak measurements of the same observable. We experimentally
demonstrate the method by directly measuring the polarization state
of an ensemble of photons while simultaneously reading out both the
real and imaginary parts of the weak value for each photon.

We begin by reviewing the theory behind weak values. We then theoretically
introduce the two above mentioned general methods to concurrently
measure the real and imaginary parts of the weak value. Next, we describe
an experiment in which we directly measure the photon polarization
state.

\section{Theoretical Description of Weak Measurement}

The concept of weak measurement is best introduced within von Neumann's
model of quantum measurement. Arguably, any type of measurement can
be described with it \cite{Wiseman2010}. The key feature of this
model that distinguishes it from the standard treatment of measurement
in quantum mechanics is that \textit{both} the measured system $S$
and the `pointer' $P$ of measurement apparatus are treated quantum
mechanically. That is, both have quantum states. The measured system
begins in an arbitrary superposition state $|\psi\rangle_{S}=\sum_{n}c_{n}|a_{n}\rangle_{S}$,
where $|a_{n}\rangle_{S}$ is an eigenstate of the observable $\hat{A}$
that is to be measured, with eigenvalue $a_{n}$. The measurement
apparatus incorporates a pointer that will indicate the result of
the measurement. The pointer is in state $|\xi\rangle_{P}$. Initially,
$|\xi\rangle_{P}=|\bar{q}=0\rangle_{P}$, a state with standard deviation
$\sigma$ and center $\bar{q}$ in some variable $q$ (e.g., position).
Together, the pointer and measured system compose the total system
$T$, which has state $|\Theta\rangle_{T}=|\psi\rangle_{S}\otimes|\bar{q}=0\rangle_{P}$
initially.

To perform the measurement, the system observable is coupled to the
conjugate pointer variable $k$ (e.g., momentum) by the von Neumann
measurement interaction
\begin{equation}
\hat{U}=\mathrm{exp}(-i\delta\hat{A}\otimes\hat{k}),\label{eq:unitary}
\end{equation}
where $\delta$ is the coupling strength. The action of unitary $\hat{U}$
is as follows. The pointer will be shifted by $\Delta q=\delta a_{n}$
for state $|a_{n}\rangle_{S}$ in $|\psi\rangle_{S}$. Thus, the final
state of the total system is given by $|\Theta'\rangle_{T}=\sum_{n}c_{n}|a_{n}\rangle_{S}|q=\delta a_{n}\rangle_{P}$,
which is now entangled between $S$ and $P$. Lastly, in the `read out' step of
the model, one measures $\hat{q}$ on the pointer to read out
the measurement of $\hat{A}$. If the shift is large compared to the
initial spread of the pointer, $\delta\gg\sigma$, then the outcome
of the measurement can be read out unambiguously in a single trial
with minimal error. However, given outcome $a_{n}$, the system will
then be left in a single state $|a_{n}\rangle_{S}$, destroying the
initial superposition (i.e., `collapse'). This is a standard (i.e.
`strong') measurement.

The opposite limit, $\delta\ll\sigma$, defines the regime of weak
measurement. In this limit, the shifted pointer states in $|\Theta'\rangle_{T}$
overlap in $q$, making the result of the measurement ambiguous in
any given trial. However, since the the measured system is now only
minimally entangled with the pointer, reading out $\hat{A}$ by measuring
$\hat{q}$ now only minimally disturbs the initial measured-system
state $|\psi\rangle_{S}$. Subsequent measurement will now reveal
additional information about that state. 
Consider a subsequent projective
measurement onto state $|\varphi\rangle_{S}$. In the sub-ensemble
of trials that have been successfully projected onto $|\varphi\rangle_{S}$
(i.e., `post-selection'), the pointer will on average be shifted by
an amount proportional to the `weak value' \cite{aharonov1988result}:
\begin{equation}
\langle\hat{A}\rangle_{W}=\frac{\langle\varphi|\hat{A|}\psi\rangle}{\langle\varphi|\psi\rangle}.\label{eq:weak_value}
\end{equation}
For this to be valid, the pointer shift, $\delta\langle\hat{A}\rangle_{W}$
(i.e., the `signal'), must be much smaller than the pointer spread
$\sigma$ (i.e., the `noise') \cite{Duck1989}. In other words, for
a single trial the signal to noise ratio in a weak measurement is
small. Nonetheless, by repeating the weak measurement in a large number
of trials one can reduce the effect of noise by averaging. The average
result is the weak value.

Unlike standard expectation values, the weak value can be complex
valued. The real and imaginary parts of the weak value manifest as
shifts in the two conjugate pointer variables $q$ and $k$: 
\begin{equation}
\langle\hat{A}\rangle_{W}=\frac{1}{\delta}\left(\left\langle \hat{q}\right\rangle +i4\sigma^{2}\left\langle \hat{k}\right\rangle \right),\label{eq:weak_value_one_pointer}
\end{equation}
where the expectation values on the right-hand side are taken for
the final pointer state.

The heart of the problem of determining $\mathrm{Re}\langle\hat{A}\rangle_{W}$
and $\mathrm{Im}\langle\hat{A}\rangle_{W}$ simultaneously is that
$\hat{q}$ and $\hat{k}$ do not commute, and, thus, can not be measured
at the same time. Instead, past experiments have measured $\mathrm{Re}\langle\hat{A}\rangle_{W}$
and $\mathrm{Im}\langle\hat{A}\rangle_{W}$ on separate sub-ensembles
of the measured system. This was achieved with one of two methods,
which we call Method A and Method B (see Fig. 1).
In Method A, the ensemble is divided in two. Each sub-ensemble is
then sent through the von Neumann interaction (i.e., Eq. \ref{eq:unitary})
and to the subsequent strong projective measurement. In the final
step, the reading out of the weak measurement, only one of the pointer
variables, $q$ or $k$, is measured for each sub-ensemble. An example
of this strategy is the original direct measurement of the wavefunction
experiment \cite{Lundeen2011}. There, the photon's polarization was
used as a pointer. The two conjugate pointer observables were $\hat{\sigma}_{x}$
and $\hat{\sigma}_{y}$, the polarization Pauli matrices. One Pauli
matrix or the other was chosen to be measured by setting the angle
of a waveplate. Thus, $\mathrm{Re}\langle\hat{A}\rangle_{W}$ and
$\mathrm{Im}\langle\hat{A}\rangle_{W}$ were determined separately
for two sub-ensembles of photons delineated by time.

In Method B, the ensemble is divided \textit{after} the von Neumann
interaction and subsequent strong measurement. An example of this
is in another direct measurement experiment \cite{Salvail2013}. In
it, the photon's polarization was the measured system and its transverse
spatial mode was used as the pointer. Just before the weak-measurement
read out, a beamsplitter divided the ensemble of photons in two. For
one sub-ensemble the transverse momentum was measured. For the other
sub-ensemble the transverse photon position was measured. In this
way, in each trial either the pointer momentum or position were determined
depending on which way the photon exited the beamsplitter. Consequently,
the real and imaginary weak values were determined with distinct sub-ensembles.

\begin{figure}[h!]
\centering \includegraphics[width=\columnwidth]{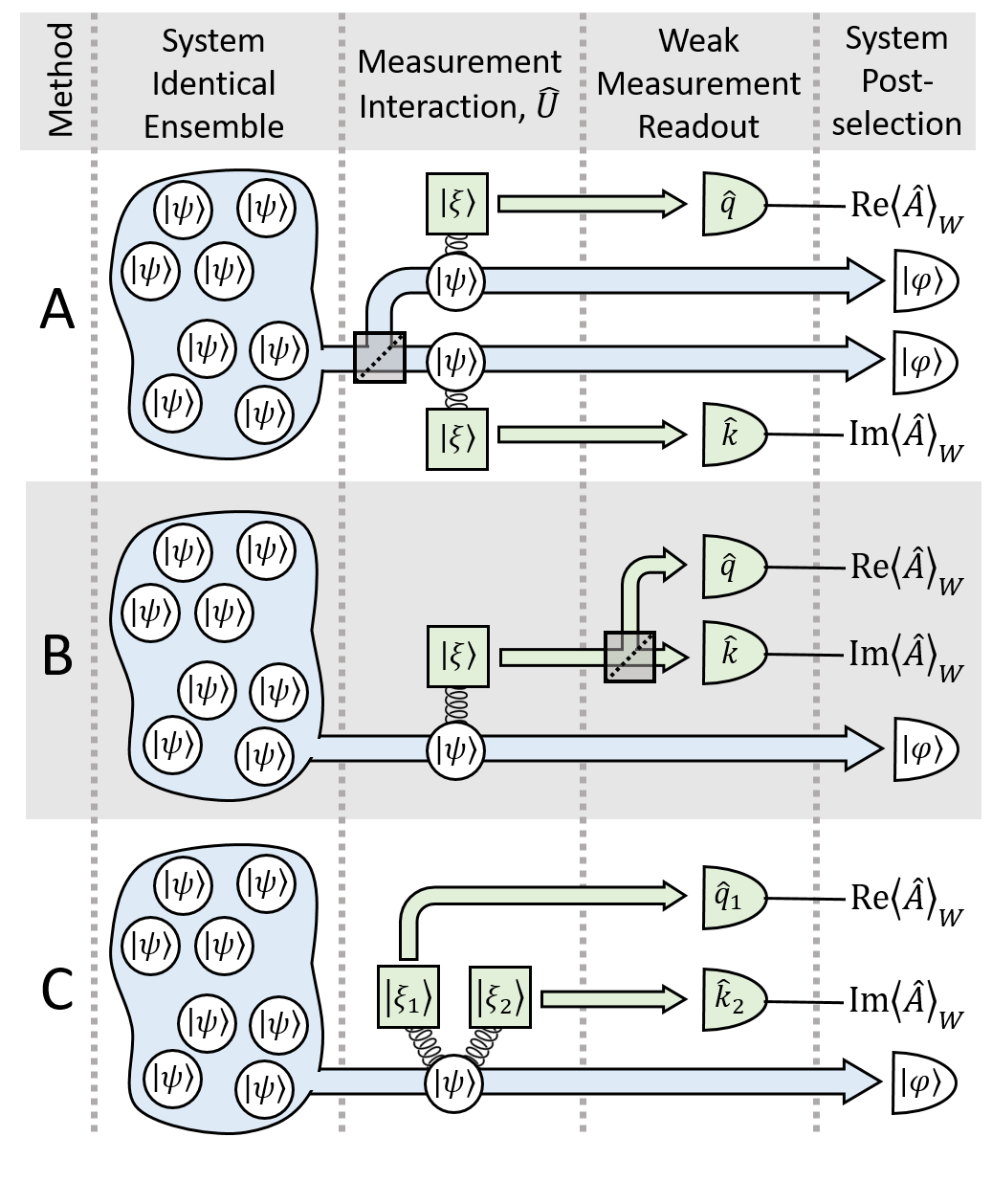}
\caption{Schemes for measuring the real and imaginary parts of the weak value,
$\mathrm{Re}\langle\hat{A}\rangle_{W}$ and $\mathrm{Im}\langle\hat{A}\rangle_{W}$.
In Method A, an identical ensemble of system states $|\psi\rangle_{S}$
is divided into two before the von Neumann measurement interaction
$\hat{U}$ couples system observable $\hat{A}$ to the pointer $P$,
initially in state $|\xi\rangle_{P}$. One sub-ensemble is then used
to read out $\mathrm{Re}\langle\hat{A}\rangle_{W}$ from the pointer.
The other sub-ensemble is used to read out $\mathrm{Im}\langle\hat{A}\rangle_{W}$.
Method B is similar except that the division into two sub-ensembles
happens after the measurement interaction. In Method C, two pointers
are coupled to the system in two measurement interactions. One pointer
is used to read out $\mathrm{Re}\langle\hat{A}\rangle_{W}$ while
the other is used read out $\mathrm{Im}\langle\hat{A}\rangle_{W}$.
Only in Method C does each member of the identical ensemble contribute
to the read out of the full complex weak.}
\end{figure}

\section{A General Method to Measure the Full Complex Weak Value}

We now introduce Method C (Fig. 1), which allows one to determine
the real and imaginary parts of the weak value simultaneously. This
was first achieved in Ref. \cite{Kobayashi2014} by using an orbital
angular momentum pointer, which had a ring-like transverse probability
distribution. Shifts of the center of the ring along two orthogonal
directions gave the two parts of the weak value. While in Ref. \cite{Kobayashi2014}
this is framed as an effect that relies on orbital angular momentum,
we will show that the key to the simultaneous determination is the
use of the two spatial directions as independent pointers.

Since weak measurements minimize disturbance to the system, multiple
weak measurements can be performed in sequence without altering their
individual results, i.e. weak values. Consider, two weak measurements
but of the same system observable $\hat{A}$. A pointer $P1$ is used
for the first von Neumann interaction and, then, another pointer $P2$,
is used for the second von Neumann interaction (Eq. \ref{eq:unitary})
giving a total evolution of
\begin{eqnarray}
\hat{U} & = & \mathrm{exp}\left(-i\delta_{1}\hat{A}\hat{k}_{1}\right)\mathrm{exp}\left(-i\delta_{2}\hat{A}\hat{k}_{2}\right)\label{eq:unitary_two_pointers}\\
 & = & \mathrm{exp}\left(-i\hat{A}\left(\delta_{1}\hat{k}_{1}+\delta_{2}\hat{k}_{2}\right)\right),
\end{eqnarray}
where $\hat{k}_{j}$ is observable $\hat{k}$ for the $j$-th pointer
(and likewise for $\hat{q}$). While $\hat{q}$ and $\hat{k}$ for
an individual pointer do not commute, $\hat{q}_{1}$ and $\hat{k}_{2}$
do commute. Consequently, the weak value can now be determined by
measuring $\hat{q}_{1}$ for the first pointer and simultaneously
measuring $\hat{k}_{2}$ for the second pointer: 
\begin{equation}
\langle\hat{A}\rangle_{W}=\frac{1}{\delta_{1}}\left\langle \hat{q}_{1}\right\rangle +i\frac{4\sigma_{2}^{2}}{\delta_{2}}\left\langle \hat{k}_{2}\right\rangle .\label{eq:weak_value_two_pointer}
\end{equation}
Note that the two pointers need not be the same type of quantum systems.
They could be an electron and photon, for example. As well, the pointer
degrees of freedom, $q_{1}$ and $q_{2}$, might be distinctly different,
e.g. spin and position. This procedure, based on a sequence of two
weak measurements of $\hat{A}$, constitutes our method to simultaneously
measure both parts of the weak value. A given trial will contribute
to the average for both $\mathrm{Re}\langle\hat{A}\rangle_{W}$ and
$\mathrm{Im}\langle\hat{A}\rangle_{W}$.

\section{Experiment to Simultaneously Measure The Full Weak Value }
\begin{figure*}[htbp]
\centering \includegraphics[width=2\columnwidth]{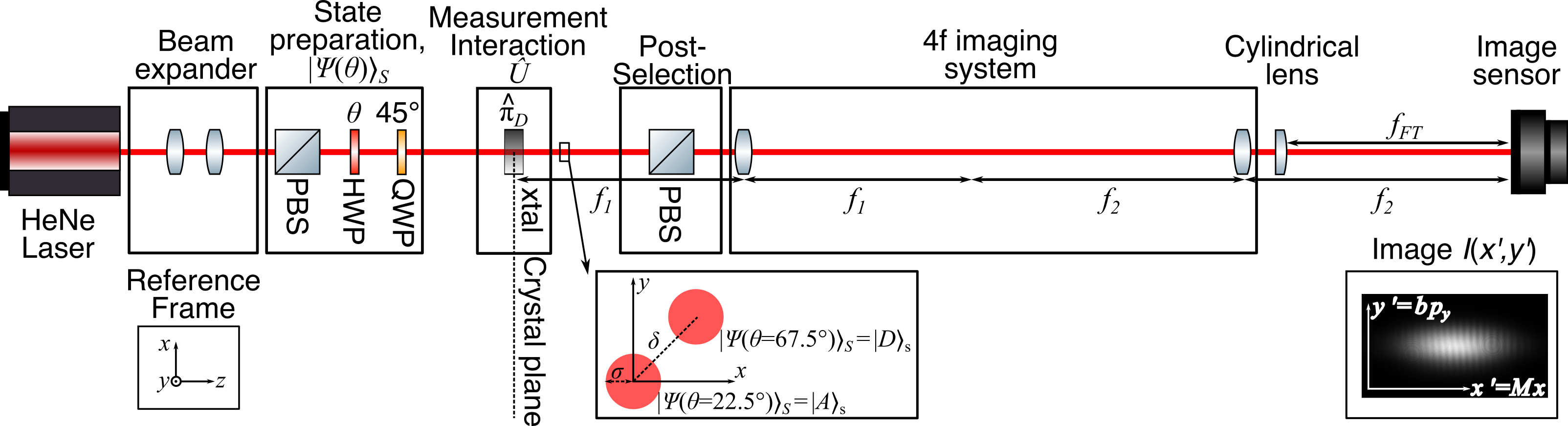}
\caption{Experimental setup for simultaneous read out of the real and imaginary
parts of the weak value. This setup implements Method C in Fig. 1.
To create the input polarization state $|\psi(\theta)\rangle_{S}$
(the state of the measured system) we use a polarizing beamsplitter
(PBS), half-waveplate (HWP) at angle $\theta$, and quarter-waveplate
(QWP). The pointers are the two transverse degrees of freedom of the
photon, $x$ and $y$. A beam expander sets the width $\sigma$ of
the pointer states. The walk-off crystal (xtal) displaces the two-dimensional
pointer along the diagonal direction by $\delta$ if the system is
diagonally polarized, $|D\rangle_{S}$. This implements the measurement
interaction by coupling the pointer to the observable $\hat{\pi}_{D}$.
A 4f lens arrangement images the crystal plane onto an image sensor.
The cylindrical lens is used to Fourier transform the spatial distribution
solely along the $y$ axis. The image sensor records the average photon
number at a pixel index $(n_{x'},n_{y'})$. This index is proportional
to $(x,p_{y})$. By evaluating the pointer shifts, $\langle\hat{x}\rangle$
and $\langle\hat{p_{y}}\rangle$, we find $\mathrm{Re}\langle\hat{\pi}_{D}\rangle_{W}$
and $\mathrm{Im}\langle\hat{\pi}_{D}\rangle_{W}$ via Eq. \ref{eq:weak_value_two_pointer}.}
\end{figure*}

We demonstrate Method C by directly measuring the polarization state
of a photon. We describe the direct measurement concept in the Appendix.
Since photons typically do not interact strongly with other quantum
systems, instead of using an external system as a pointer, we use
two photonic degrees of freedom as the system and pointer, polarization
and transverse mode. This allows us to use linear optics to implement
the von Neumann interaction, which will couple a polarization observable
to the transverse position of the photon.

For a two-dimensional pointer, such as transverse position, Method
C can be significantly simplified. A photon traveling along $z$ has
transverse position $(x,y),$ so that $q_{1}=x$ and $q_{2}=y$. Consider
if the two couplings are equal, $\delta_{1}=\delta_{2}\equiv\sqrt{2}\delta$.
With this, the two-pointer unitary in Eq. \ref{eq:unitary_two_pointers}
reverts to the standard von Neumann interaction (Eq. \ref{eq:unitary})
with a single pointer and a single pointer degree of freedom: 
\begin{equation}
\hat{U}=\mathrm{exp}\left(-i\delta\hat{A}\hat{p}_{D}/\hbar\right),\label{eq:unitary_diag}
\end{equation}
where $\hat{x}_{D}=(\hat{x}+\hat{y})/\sqrt{2}$ is the photon position
along the diagonal direction $D$ (it is the $\sqrt{2}$ in $\hat{x}_{D}$
that necessitates the $\sqrt{2}$ in $\delta$). While, we now only
require the standard von-Neumann single-pointer unitary interaction
(Eq. \ref{eq:unitary}), the final pointer readout must still be two-dimensional.
That is both $x$ and the transverse momentum $p_{y}$ (i.e., along
$y$) must be measured in order to simultaneously evaluate both $\mathrm{Re}\langle\hat{A}\rangle_{W}$
and $\mathrm{Im}\langle\hat{A}\rangle_{W}$ according to Eq. \ref{eq:weak_value_two_pointer}.

The experimental setup is shown in Fig. 2. Our ensemble of photons
is produced by a Helium-Neon laser with a wavelength of $\lambda=633$
nm. A polarizing beamsplitter (PBS) sets the photon polarization state
to horizontal $|H\rangle_{S}$. The spatial distribution is Gaussian
in both the $x$ and $y$ directions and is set to have a $1/e^{2}$
half-width of $\sigma=\sigma_{x}=\sigma_{y}=306\pm2$ $\upmu \mathrm{m}$
with a beam expander. We use a half-waveplate (HWP) with its optical
axis at an angle $\theta$ from the horizontal followed by a quarter-waveplate
(QWP) with its axis $-45^{\circ}$ from the horizontal to produce
an input polarization state, 
\begin{eqnarray}
|\psi(\theta)\rangle_{S} & = & c_{D}|D\rangle_{S}+c_{A}|A\rangle_{S}\label{eq:actual_input}\\
 & = & \sin(2\theta-\pi/4)|D\rangle_{S}-i\cos(2\theta-\pi/4)|A\rangle_{S},\nonumber 
\end{eqnarray}
where $|D\rangle$ and $|A\rangle$ are the diagonal and anti-diagonal polarization states, respectively. In order to test our method with a range of system states, we vary $|\psi(\theta)\rangle_{S}$ by rotating the angle $\theta$ of the HWP.

We implement the von Neumann interaction by using a birefringent crystal
(Beta Barium Borate) to couple the polarization of the photon to its
spatial distribution. A photon with a polarization along the crystal
optical axis will be transversely displaced along the direction of
the axis, whereas a photon with the orthogonal polarization will not.
We use a single crystal (xtal) to displace the position $x_D$ of photons with polarization
$\left|D\right\rangle_S $ by
$\delta=163\pm2$ $\upmu$m. Since $ \delta < \sigma$, this interaction is in the weak measurement regime. The weakly measured observable
is $\hat{\pi}_{D}\equiv\left|D\right\rangle \left\langle D\right|_{S}$.
While, in our case, the displacement and polarization are collinear,
by sandwiching such a `walk-off crystal' between waveplates, any polarization
projector can be measured.

Following the walk-off crystal, we project the measured system $|\psi\rangle_{S}$
onto the horizontal polarization state $|H\rangle_{S}$ with a polarizing
beamsplitter. The transmitted photons are then sent through a 4f lens
system (spherical lenses, focal lengths $f_{1}=1$ m and $f_{2}=1.2$
m) that images the crystal plane onto an image sensor (a CMOS sensor
with resolution $2560\times1920$ and a pixel pitch of 2.2 $\upmu \mathrm{m}$
$\times$ 2.2 $\upmu \mathrm{m}$). The final photon position $x'$ on the sensor
is given by $x'=Mx$, where magnification $M=f_{2}/f_{1}=1.2$. The
use of a 4f system creates room between the image sensor and the xtal
for a long focal length cylindrical lens ($f_{FT}=1$ m, curved along
the $y$-direction) to be placed one focal length before the sensor.
The cylindrical lens performs a Fourier transform such that the final
position is proportional to the initial transverse momentum, $y'=bp_{y}$, where
$b=\lambda f_{FT}/2\pi\hbar$. The large value of $f_{FT}$ ensures
the distribution will cover many pixels in the $y'$ direction.

To read out the result of the weak measurement we must determine the average shift of the pointer along $x'$ and $y'$. Thus, we record the average number of photons (i.e., the
intensity) detected at a given position, $I(x',y')$. From this $\mathrm{Re}\langle\hat{\pi}_{D}\rangle_{W}$
and $\mathrm{Im}\langle\hat{\pi}_{D}\rangle_{W}$ can be calculated
by taking $x=x'/M$ and $p_{y}=y'/b$ in Eq. \ref{eq:weak_value_two_pointer}.
However, we do not do this. Because it is more direct, a more accurate
method is to calculate the expectation values in terms of pixel index
$(n_{x'},n_{y'})$ rather than position:
\begin{equation}
\langle\hat{A}\rangle_{W}=\frac{1}{\delta_{x'}}\left(\left\langle \hat{n}_{x'}\right\rangle +i\frac{\sigma_{x'}}{\sigma_{y'}}\left\langle \hat{n}_{y'}\right\rangle \right).\label{eq:pixel_weak_value}
\end{equation}
To arrive at this expression, $M$ has been expressed in terms of
$\delta_{x'}=62.8\pm0.9$ and $\sigma_{x'}=167\pm1$, which are $\delta$
and $\sigma$ in units of pixels. In addition, $b$ has been expressed
in terms of $\sigma_{y'}=62.9\pm0.4$ in units of pixels, the pointer
width along the $y'$ pixel direction. See Ref. \cite{Thekkadath2016}
for the details of this method. We vary input system state $|\psi(\theta)\rangle_{S}$
over range $\theta=0$ to $90^{\circ}$ in $3^{\circ}$ steps.   For each step
three images were collected and averaged and then used to determine pointer shifts $\left\langle \hat{n}_{x'}\right\rangle$ and $\left\langle \hat{n}_{y'}\right\rangle$. The full range was stepped through seven times. These seven trials were used to determine the mean pointer shifts and their standard error. Based on these, in the next section, we present the measured weak values.

\section{Results}

In Fig. 3. a) and b), we respectively plot the $\mathrm{Re}\langle\hat{\pi}_{D}\rangle_{W}$
and $\mathrm{Im}\langle\hat{\pi}_{D}\rangle_{W}$ points experimentally
determined using Method C. The solid line plotted in each panel of
Fig. 3 is the corresponding theoretical prediction:
\begin{eqnarray}
\langle\hat{\pi}_{D}\rangle_{W} & = & \frac{c_{D}}{c_{D}+c_{A}}\label{eq:theory_weak_value}\\
 & = & \frac{1}{2}\left(1+\cos\left(4\theta+\frac{\pi}{2}\right)\right)-\frac{i}{2}\sin\left(4\theta+\frac{\pi}{2}\right),\nonumber
\end{eqnarray}
which is found from Eq. \ref{eq:actual_input} and Eq. \ref{eq:weak_value_two_pointer}.
This weak measurement directly measures the amplitudes of $|\psi\rangle_{S}$.
Note that the $c_{A}$ amplitude can be eliminated or determined through
normalization $|c_{A}|^{2}=1-|c_{D}|^{2}.$ The phases of the amplitudes
are determined up to a global phase that changes with the input state
$|\psi\rangle_{S}$.

\begin{figure}[h!]
\centering \includegraphics[width=\columnwidth]{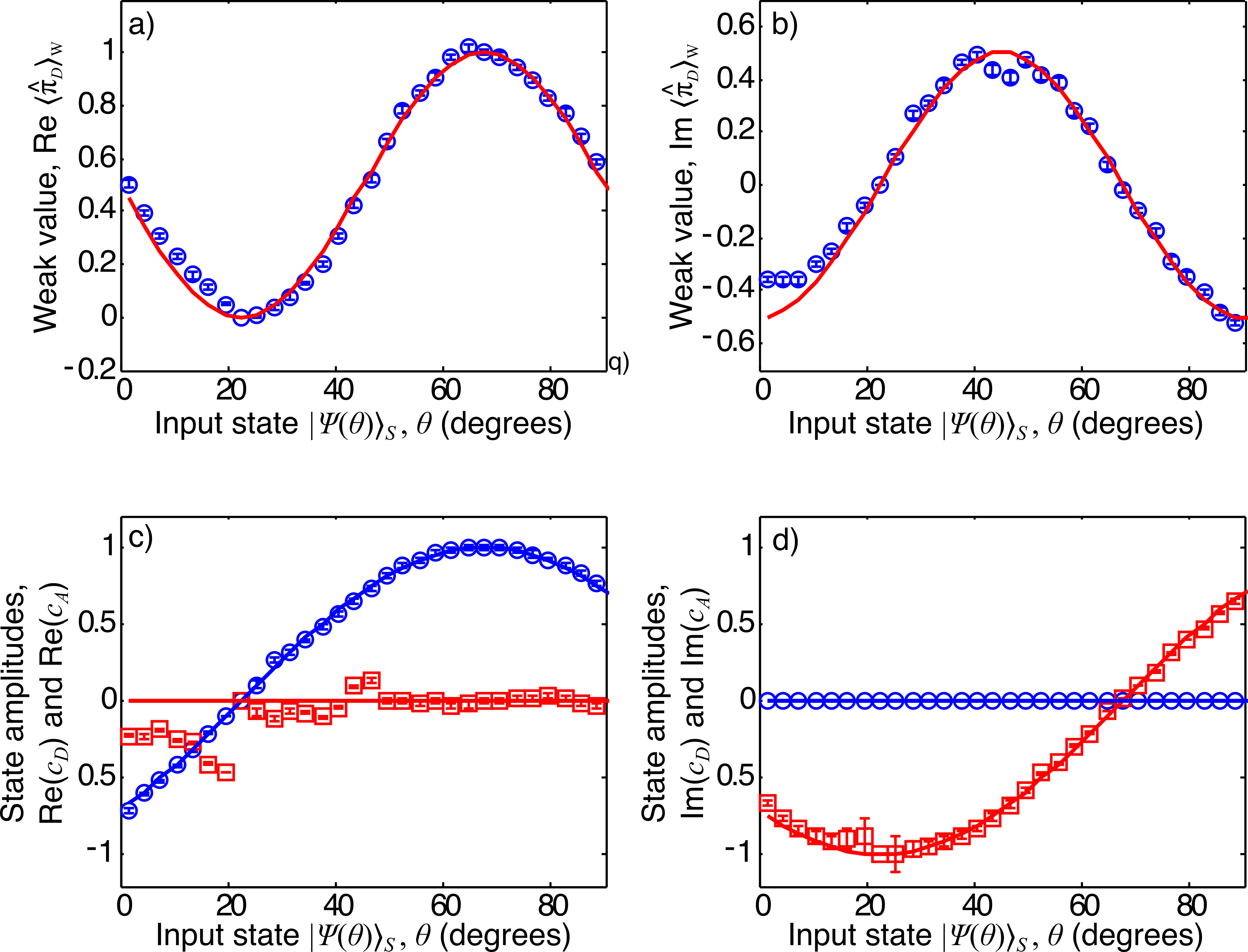}
\caption{Real and imaginary parts of the measured weak value. In a) and b),
we plot the results of a weak measurement of $\hat{\pi}_{D}$ using
Method C in Fig. 1. The weak value $\langle\hat{\pi}_{D}\rangle_{W}$
is found from the measured shifts along $x'$ and $y'$ according
to Eq. \ref{eq:pixel_weak_value} (blue circles). For system state
$|\psi(\theta)\rangle_{S}=c_{D}|D\rangle_{S}+c_{A}|A\rangle_{S}$
the predicted weak value is given by the Eq. \ref{eq:theory_weak_value}
(red line). In c) and d), we use the fact that $\langle\hat{\pi}_{D}\rangle_{W}\varpropto c_{D}$
to directly measure the system state (see the Appendix for details).
We plot measured system state amplitudes, $c_{D}$ (blue circles)
and $c_{A}$ (red squares). The nominal amplitudes are indicated by
the solid lines given by Eq. \ref{eq:actual_input}. The error bars
indicate the standard error found from repeated trials.}
\end{figure}

The data points closely follow the theoretical curve, indicating that
Method C works. However, they do not agree with theory to within error.
We attribute this to systematic errors such as offsets from the nominal
birefringent retardance of the waveplates, the axis direction of the
crystals, and movement of the beam with waveplate rotation.

\section{Discussion and Conclusion}

Since the introduction of weak values $\langle\hat{A}\rangle_{W}$
there has been confusion and disagreement about what they represent
\cite{Leggett1989,Peres1989,Aharonov1989,Ferrie2014,Hofmann2014a,Brodutch2015,Ferrie2015,vaidman2017weak}.
Should they be interpreted as the value of the weakly measured parameter
\cite{Vaidman2017} (even though it can lie outside the range of the
observable's eigenvalues \cite{Aharonov1988}), as the average value
of that parameter \cite{Mir2007,Kocsis2011}, as a real deterministic
property of the measured system \cite{Aharonov1991,Vaidman1996a,Hofmann2012a},
or perhaps not regarded as a measurement result at all? This confusion
has been compounded by the fact that the weak value is generally complex,
in contrast to the standard quantum expectation value. Indeed, in
some papers the imaginary part of the weak value is not considered
part of the result of the weak measurement at all \cite{Wiseman2003}.

There are arguments for and against considering $\mathrm{Im}\langle\hat{A}\rangle_{W}$
as part of the result of the weak measurement. An argument in favour of this
is that both $\mathrm{Re}\langle\hat{A}\rangle_{W}$ and $\mathrm{Im}\langle\hat{A}\rangle_{W}$
have clear physical manifestations: as shifts in two conjugate variables
of the measurement apparatus pointer, e.g., position and momentum,
respectively. An ``against'' argument is that the appearance of
an imaginary component of the average result of a measurement is completely
unexpected and, in the context of probability theory, nonsensical.
Another ``against'' argument is that the momentum shift is usually
much smaller than the position shift and thus, should be considered
a side-effect of the measurement (i.e., reverse `back-action' \cite{Steinberg1995b}).

This confusion about the complex nature of the weak value is compounded
by the fact that $\mathrm{Re}\langle\hat{A}\rangle_{W}$ and $\mathrm{Im}\langle\hat{A}\rangle_{W}$
were not measured simultaneously in experiments. That is, since the
two pointer variables are conjugate, the two shifts could not be determined
at the same time. This is particularly consequential when considering
the weak-measurement-based concept of direct measurement of the wavefunction.
There, the directness is partly a reflection of the full weak value
appearing on the measurement apparatus in a straight-forward manner.
If the real and imaginary parts do not appear simultaneously then
it puts this directness in question.

In this paper, we introduced and experimentally demonstrated a general
method to simultaneously measure both the real and imaginary components
of the weak value. The method uses a separate pointer
and von Neumann measurement interaction for each weak value component.
We simplified the method in the case of an inherently two-dimensional
system, such as transverse position, so that only one measurement
interaction is required. In the method, each and every trial contributes
to both the $\mathrm{Re}\langle\hat{A}\rangle_{W}$ and $\mathrm{Im}\langle\hat{A}\rangle_{W}$
averages. Thus, two pointer shifts can manifest themselves at the same time,
giving the real and imaginary parts of the wavefunction in a direct
measurement. In summary, this paper has provided support to the notion
that the full complex weak value should be considered the average
result of a weak measurement, and, in that sense, a fundamental property
of the measured system.

\section*{Acknowledgments}

This work was supported by the Canada Research Chairs (CRC) Program,
the Canada First Research Excellence Fund (CFREF), and the Natural
Sciences and Engineering Research Council (NSERC). Hariri was supported
by a Mitacs Globalink Research Internship.

\section*{Appendix: Direct Measurement Of The Quantum State}

Our experimental demonstration performs a direct measurement of quantum
state. Below we describe what is meant by a direct measurement. Consider
if one want to measure the amplitudes of an arbitrary polarization
state of a photon expressed in the $A/D$ basis, 
\begin{equation}
|\psi\rangle_{S}=c_{D}|D\rangle_{S}+c_{A}|A\rangle_{S},\label{eq:polstate}
\end{equation}
where $c_{D}=\left\langle D|\psi\right\rangle $ and $c_{A}=\left\langle A|\psi\right\rangle $.
We define $|H\rangle,|V\rangle,|D\rangle,$ and $|A\rangle,$ as the
horizontal, vertical, diagonal, and anti-diagonal polarization states,
where $\left|D\right\rangle =(\left|H\right\rangle +\left|V\right\rangle )/\sqrt{2}$.
The concept for direct measurement was introduced in Ref. \cite{lundeen2011direct}.
In it, a weak measurement of a variable is followed by a strong measurement
of a complementary variable. The weak value is proportional to the
quantum state. For polarization, this entails weakly measuring $\hat{\pi}_{J}\equiv\left|J\right\rangle \left\langle J\right|$
for $J=A,D$ and then strongly projecting the measured system on $|H\rangle$.
A successful projection defines a sub-ensemble of trials in which
the average result of the weak measurement, the weak value, is 
\begin{equation}
\langle\hat{\pi}_{J}\rangle_{W}=\frac{\langle H|\hat{\pi}_{J}|\psi\rangle}{\langle H|\psi\rangle}=\sqrt{N}\left\langle J|\psi\right\rangle ,\label{eq:direct_weak}
\end{equation}
where $\sqrt{N}$ is a constant, independent of $J$.

In terms of this weak value, the quantum state is given by, 
\begin{eqnarray}
|\psi\rangle_{S} & = & \frac{1}{\sqrt{N}}\left(\langle\hat{\pi}_{D}\rangle_{W}|D\rangle_{S}+\langle\hat{\pi}_{A}\rangle_{W}|A\rangle_{S}\right)\\
 & = & \frac{1}{\sqrt{N}}\left(\langle\hat{\pi}_{D}\rangle_{W}|D\rangle_{S}+\left(1-\langle\hat{\pi}_{D}\rangle_{W}\right)|A\rangle_{S}\right).\label{eq:direct}
\end{eqnarray}
The second line is a simplification using $\hat{I}=\hat{\pi}_{D}+\hat{\pi}_{A}$.
Normalization fixes $N=(|\langle\hat{\pi}_{D}\rangle_{W}|^{2}+|1-\langle\hat{\pi}_{D}\rangle_{W}|^{2})$.
Typically, one would also fix the global phase, which otherwise would
vary with $|\psi\rangle_{S}$. We do this by setting $c_{D}=|c_{D}|\exp(i\phi_{D})$
to be real always. Summarizing, for polarization we need only weakly
measure $\hat{\pi}_{D}$ to directly measure the quantum state. This
procedure was demonstrated in Ref. \cite{Salvail2013}, but as discussed
in the main body of the paper, a beamsplitter was used to randomly
measure either $x$ or $p_{x}$ for a given photon. In contrast, in
our method, each member of our photon ensemble will contribute to
both both the $\mathrm{Re}\langle\hat{A}\rangle_{W}$ and $\mathrm{Im}\langle\hat{A}\rangle_{W}$
averages.

\bibliographystyle{apsrev4-1}
\bibliography{mybib}

\end{document}